\documentclass[12pt]{emulateapj}
\usepackage[english]{babel}
\usepackage{amsmath}
\usepackage{amssymb}
\usepackage[dvips]{color}
\usepackage{natbib}

\newcommand{\sourO}{IGR J17480--2446}
\newcommand{\sour}{J17480}

\newcommand{\nus}{\nu_{\rm{s}}}
\newcommand{\nusten}{\nu_{\rm{s},10}}
\newcommand{\nur}{\nu_{\rm{r}}}
\newcommand{\nuo}{\nu_{\rm{o}}}
\newcommand{\Ro}{R_{\rm{R}}}

\begin{document}

\title{Implications of burst oscillations from the slowly rotating 
accreting pulsar\\ \sourO{} in the globular cluster Terzan 5}
\author{Y. Cavecchi\altaffilmark{1,2}, A. Patruno\altaffilmark{1},
  B. Haskell\altaffilmark{1},  A.L. Watts\altaffilmark{1}, Y. Levin\altaffilmark{2,3},
  M. Linares\altaffilmark{4}, D. Altamirano\altaffilmark{1}, 
R. Wijnands\altaffilmark{1} and M. van der Klis\altaffilmark{1}}
\altaffiltext{1}{Astronomical Institute ``Anton Pannekoek'', University of
  Amsterdam, Postbus 94249, 1090 GE Amsterdam, the Netherlands}
\altaffiltext{2}{Sterrewacht Leiden, University of Leiden,
Niels Bohrweg 2, NL-2333 CA Leiden, The Netherlands}
\altaffiltext{3}{School of Physics, Monash University, P.O. Box 27,
  Victoria 3800, Australia}
\altaffiltext{4}{MIT Kavli Institute for Astrophysics and Space
  Research, 70 Vassar St., 02139 Cambridge, USA}
\altaffiltext{5}{Email: Y.Cavecchi@uva.nl}

\begin{abstract}
\noindent
The recently-discovered accreting X-ray pulsar \sourO{} 
spins at a frequency of $\sim$11 Hz. We show that Type I X-ray bursts
from this source display 
oscillations at the same frequency as the stellar spin. 
\sourO{} is the first secure case of a slowly rotating neutron star
which shows Type I burst oscillations, all other sources featuring
such oscillations spin at hundreds of Hertz.  This
means that we can test
burst oscillation models in a completely different regime.  We explore the 
origin of Type I burst oscillations 
in \sourO{} and conclude that they are not caused 
by global modes in the neutron star ocean. We also show that the Coriolis
force is not able to confine an oscillation-producing hot-spot
on the stellar surface. 
The most likely scenario is that the burst oscillations are produced
by a hot-spot confined by hydromagnetic stresses.
\end{abstract}
\maketitle

\section{Introduction}
Accreting neutron stars (NSs) in low mass X-ray binaries  
show bright Type I X-ray bursts. These begin with a rapid
increase of the X-ray flux (the rise) followed by
a slow decrease (the tail) to 
the pre-burst luminosity, and last $\sim$10-100 s. These bursts are 
powered by thermonuclear runaways, which burn up
a layer of accumulated light elements on the NS surface
\citep[for a review see][]{rev-2003-2006-stro-bild-arxivpaper}.

A significant fraction of bursts display quasi-periodic modulations,
known as burst oscillations \citep[BOs,][]{art-1996-stro-etal}.  In
the discovery paper, 
\citeauthor{art-1996-stro-etal} suggested that BOs were related to the
spin frequency of the NS.  This has been confirmed in five
accretion-powered millisecond X-ray pulsars (AMXPs), where it was
found that BO frequencies are within a few percent of the spin
frequencies \citep{rev-2003-2006-stro-bild-arxivpaper}. This implies
that BOs are caused by a near-stationary temperature asymmetry which
persists in the surface layers of the star during burst.

The detailed phenomenology of BOs, however, is diverse.  In the
sources which have evidence for substantial magnetic fields (the
persistent AMXPs), for example, BOs appear at a near constant
frequency in the tail, with some fast chirps in the rise
\citep{art-2003-chak-morgan-etal,art-2003-stroh-mark-swank-zand,
  art-2010-alta-watts-etal}.  In the intermittent AMXPs, and NSs
without evidence for a dynamically important magnetic field, BOs
typically drift upwards by a few Hertz during the burst
\citep{art-2002-muno-etal,art-2009-watts-al}.  Both the origin of the
surface temperature asymmetry that causes the BOs, and 
the reason for the observed frequency drifts, remain unsolved puzzles.

One possibility is that the asymmetry is caused by global
modes (waves) that develop in the bursting ocean
\citep{art-1996-stro-lee,art-2004-hey,art-2005-cum}.  As the ocean
cools, its scale height $H$ decreases and the pattern speed, which
scales as $\sqrt{H}$, changes, leading to frequency drift.  The kind
of modes that might be excited include r-modes \citep{art-2004-hey},
g-modes \citep{art-1998-cumm-bild} or magneto-hydrodynamical modes
\citep{art-2009-heng-spit}.  To date none of these models have managed
to explain both the observed frequencies and the magnitude of the
drifts \citep[see e.g.][]{art-2005-piro-bild,art-2008-berk-levin}.

An alternative possibility is that a compact burning hot-spot develops
on the surface \citep{art-1996-stro-etal}.  The question is then how
confinement, of fuel or the flame front itself, might be achieved. For
unmagnetized stars, the Coriolis force could be an effective confining
mechanism\defcitealias{art-2002-spit-levin-ush}{SLU02}%
\citep[][hereafter SLU02]{art-2002-spit-levin-ush}. Although lifting
and expansion of the hot fluid should cause spreading of the burning
fuel, the Coriolis force would oppose such motion of the flame front
by deflecting its velocity.  This mechanism could account for the
presence of oscillations in the burst rise, although it does not
easily explain the presence of BOs in the tail or the frequency
drifts.  The Coriolis force was however attractive in that it could
explain why BOs had not been seen in any NS with spin frequency $\nus
\lesssim 245$ Hz (since for more slowly rotating stars the Coriolis
force is not dynamically relevant, see Sec. \ref{sec_coriolis}).

A strong magnetic field could also lead to confinement of fuel or
flame, with the restoring force supplied by field pressure or
stress. This mechanism is particularly 
plausible for accreting pulsars, where the existence of dynamically
important magnetic fields 
is suggested by the presence of accretion-powered pulsations (APPs).

Since there is no model that definitively explains all features of BO
phenomenology, it is important to explore the applicability of the
models under the widest possible range of conditions. Until now,
however, the only stars to show BOs were rapid rotators with $\nus
\gtrsim 245$ Hz.  This situation has changed with our discovery of
burst oscillations \citep{atel-2010-alta-etal} from the accreting
pulsar \sourO{} (hereafter \sour), which rotates an order of magnitude
more slowly at $\sim$ 11 Hz \citep{atel-2010-stroh-mark}.

\section{Observations and Data Analysis}
\label{sec_obs}

\sour{} was detected in outburst on 10 October 2010 in a Galactic
bulge scan with INTEGRAL
\citep{atel-2010-bordas-etal}. The source is located in the globular
cluster Terzan 5 and its outburst lasted for $\sim$55 days
(according to MAXI monitoring observations) before the source
became undetectable due to solar constraints. Further monitoring
observations by MAXI showed that the source had probably returned
to quiescence by January 2011. \sour{} is an 11 Hz pulsar in
a 21.3 hr orbit around a companion with
$M>0.4\,M_{\odot}$ \citep{atel-2010-stroh-mark,
art-2011-papit-etal}. Follow-up observations of \textit{RXTE}/PCA
have collected approximately 294 ks of data to 19 November 2010. We
used all \textit{RXTE}/PCA \citep{art-2006-jahoda-etal} public
observations of \sour{} taken between MJD 55482.0 and 55519.2
(October 13-November 19, 2010). For our coherent timing analysis
we used Event and GoodXenon data modes, keeping only absolute
channels 5--37 ($\sim 2 - 16$ keV), rebinned to $1/8192$ s and
barycentered with the FTOOL \texttt{faxbary} using the
position of \citet{atel-2010-pooley-etal}.

\subsection{Lightcurve and accretion rate}
\label{subsec_light}

We estimate the flux during each observation from the background
subtracted $2-16$ keV Crab normalized intensity \citep[estimated
following][]{art-2003-straat-etal} and converted it to $\textrm{erg
  cm$^{-2}$ s$^{-1}$}$ assuming a Crab spectrum for the source.  Flux
is converted into luminosity using a distance of 5.5 kpc
\citep{art-2007-orto-etal}. Luminosity $L_x$ is converted into mass
accretion rate $\dot{M}$ using $L_x=20\%\dot Mc^2$
\citep[]{book-2002-frank-king-raine-3}. The value of $\dot{M}$
obtained is approximate, since the X-ray flux is not bolometric, and
no correction has been applied for absorption, the unknown disc
inclination or gravitational redshift.  Conversion of bolometric
luminosity to mass accretion rate is further complicated by the
uncertainty in the radiative efficiency. The overall uncertainty on
$\dot{M}$ is within a factor $\sim$ 3, taking into account all these
corrections \citep{art-2007-zand-jon-markw}.  The lowest luminosity
observed gives $\dot{M}=9\times10^{-10}\rm\,M_{\odot}\,yr^{-1}$ (in
the first observation) while at the outburst peak $\dot
M=7\times10^{-9}\rm\,M_{\odot}\,yr^{-1}$, i.e., 5\% and 37\% of the
Eddington rate $\dot M_{Edd}=1.9\times10^{-8}\rm\,M_{\odot}\,yr^{-1}$
(assuming solar composition for the accreted gas).

\subsection{Bursts}

Bursts were identified visually in a 1 s lightcurve.  The beginning
and end of the bursts were defined as the points where the flux first
and last exceeded the maximum pre-burst flux in the same
observation. Typical burst length is $\sim100-200$ s.  Burst
recurrence time and burst peak-to-persistent flux ratio decrease with
increasing persistent flux until the outburst peak when the bursts
disappear and millihertz quasi-periodic oscillations
\citep{atel-2010-lina-etal} appear.  As the outburst flux decreases
again, the bursts gradually reappear with recurrence time and burst
peak-to-persistent flux ratio increasing with decreasing flux
\citep{art-2011-chako-bhatta}. Between MJD 55486 and MJD 55492 the
difference between bursts  and persistent flux fluctuations
becomes negligible and burst recurrence time is of the same order as
the duration of the bursts themselves, therefore burst rise and tail
become difficult to define. In this case we do not identify the
bursts, and for the purpose of measuring the
oscillations treat all data as persistent flux.

\begin{figure}
\centering
\includegraphics[width=0.45\textwidth]{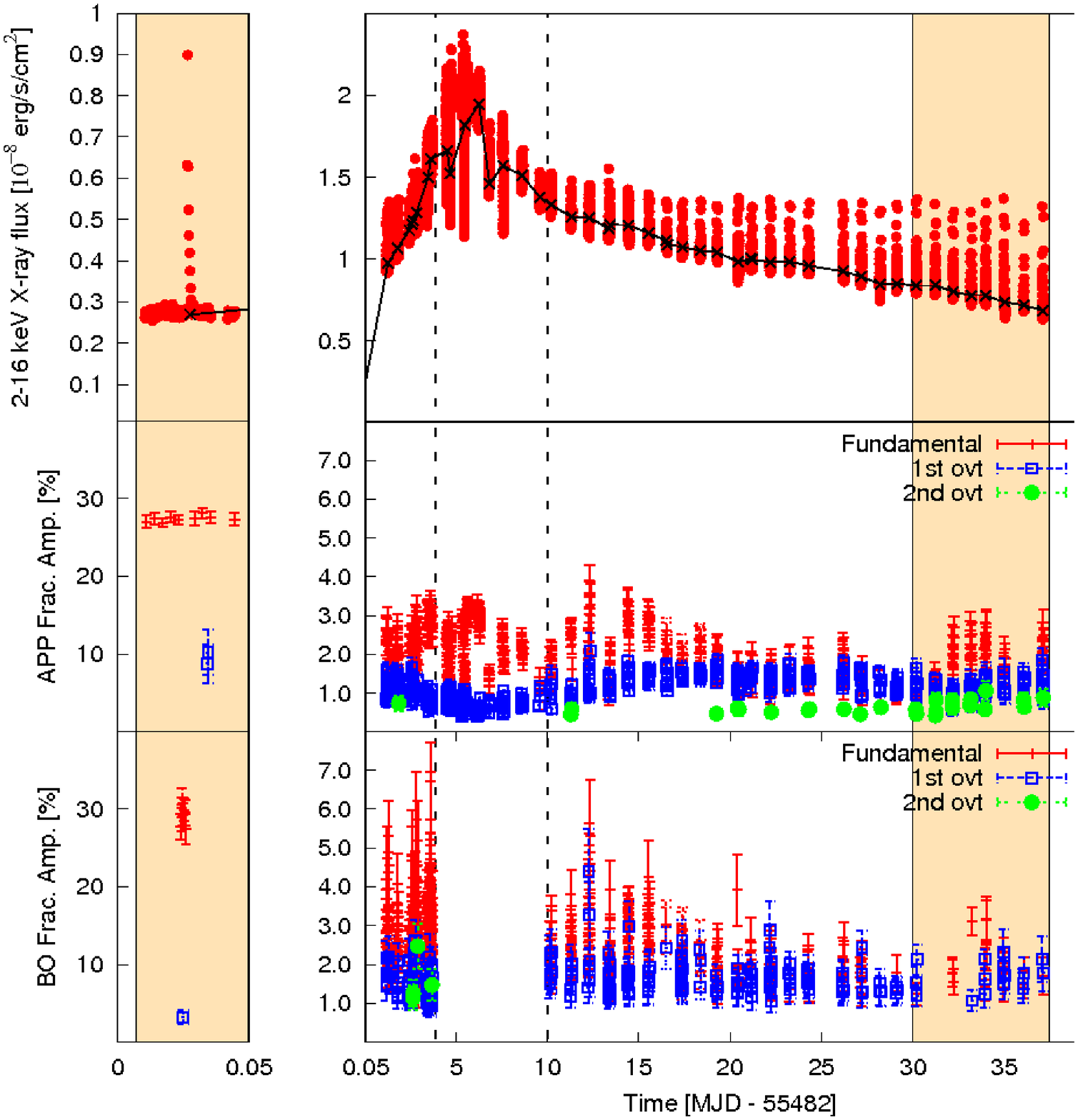}
\caption{\textbf{Top panel}: 2-16 keV X-ray lightcurve of \sour{}
  averaged over 16 s long data intervals (red dots) and average flux
  per ObsId (black crosses). Shaded regions contain the bursts with
  clear evidence of cooling
  \citep{art-2011-chako-bhatta,art-2011-manu-mhric}, note that the
  first region has a different scaling for both axes. Vertical dashed
  lines mark the region where X-ray bursts are difficult to define, so
  are excluded from the burst analysis. \textbf{Middle panel}:
  Sinusoidal fractional amplitudes of the \textbf{APP}s for the
  fundamental frequency and first two overtones. Fractional amplitude
  is $\sim28$\% in the first observation, subsequently dropping to
  $1-2\%$. The fundamental and first overtone are detected throughout
  the outburst; overtones are seen only sporadically. \textbf{Bottom
    panel}: Sinusoidal fractional amplitude of the \textbf{BO}s for
  the fundamental frequency and first two overtones. Fractional
  amplitude evolution is very similar to that of the APPs, dropping
  after the first observation from $28\%$ to $\sim 1-4\%$. }
\label{fracamp}
\end{figure}

Using this identification criterion, we found 231 bursts. As discussed
by other authors, some showed clear spectral evidence for cooling
\citep{art-2011-chako-bhatta,art-2011-motta,art-2011-manu-mhric},
were observed when the persistent luminosity was $\lesssim15\%$
Eddington and were identified conclusively as Type I (thermonuclear)
bursts (Fig. \ref{fracamp}).  The nature of the other bursts, observed
when the persistent flux was $\sim15-35\%$ Eddington, remains debated
due to the lack of clear evidence for cooling
\citep{atel-2010-gall-zand,art-2011-chako-bhatta}.

\begin{figure}
\centering
\includegraphics[width=0.45\textwidth]{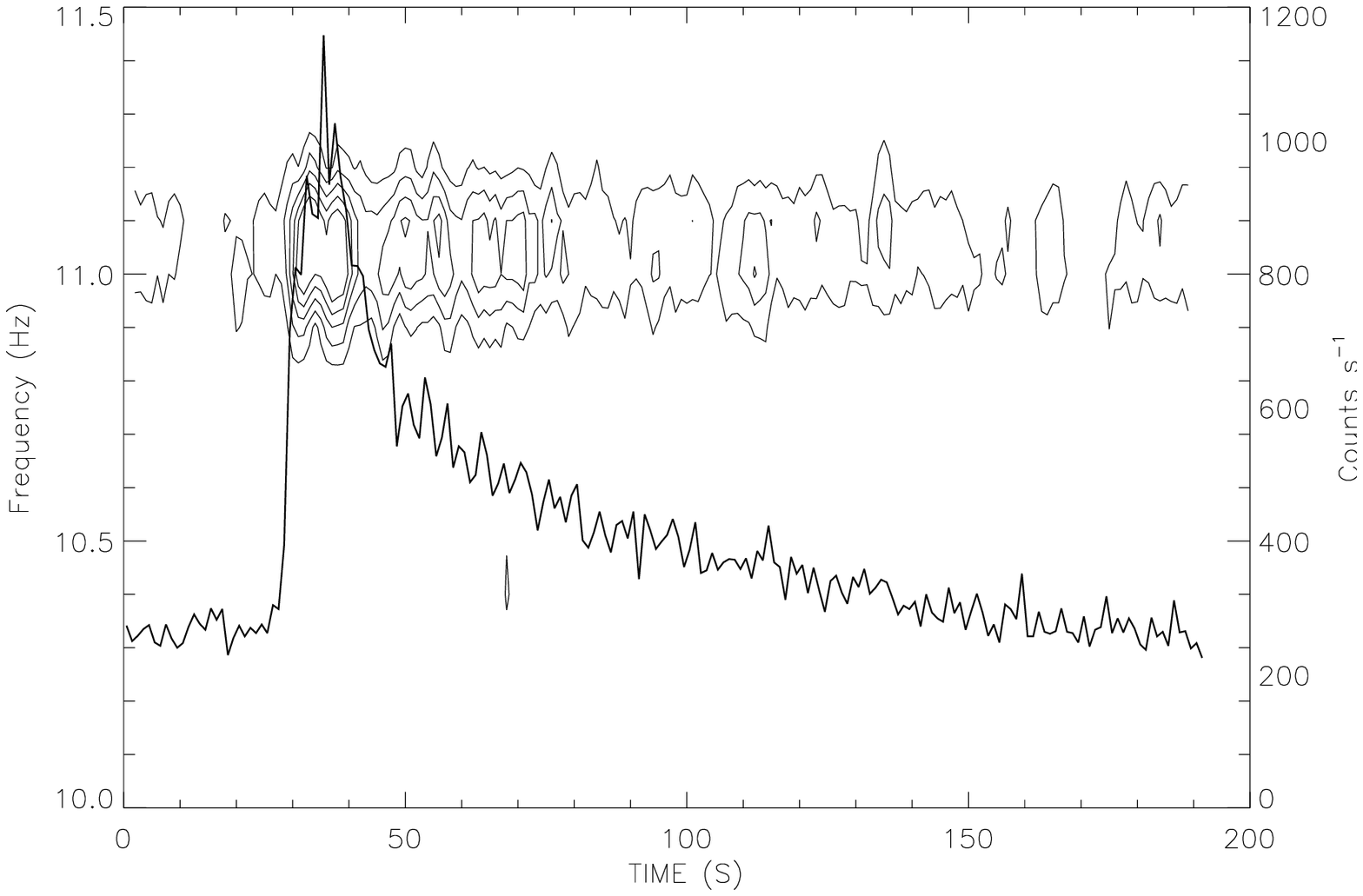}
\caption{Light curve and dynamical power spectrum for the 2010 October
  13 thermonuclear \citep[as showed by the spectral analysis
  of i.e.][]{art-2011-manu-mhric} burst, using GoodXenon data from PCU2, the
  only detector active at the time. The dynamical power spectrum uses
  overlapping 4 s bins, with new bins starting at 1 s intervals. The
  contours show Leahy normalized powers of 30, 60, 90, 120 and 150.}
\label{powerspec}
\end{figure}

\begin{table}
\centering
\caption{}
\begin{tabular}{cll}
\hline
\hline
 & $\nu$ [Hz] & $\dot\nu$ [$10^{-12}\rm\,Hz\,s^{-1}$]\\
\hline
Fundamental BOs      &11.04488532(3) &1.44(3)\\
Fundamental APPs     &11.04488540(5) &1.42(5)\\
\hline
1st Overtone BOs     &11.04488547(5) &1.33(5)\\
1st Overtone APPs    &11.04488548(10)&1.34(11)\\
\hline
2nd Overtone BOs     & ---	     & ---   \\
2nd Overtone APPs    &11.04488530(8) &1.40(6)\\
\hline
\end{tabular}
\tablecomments{All frequencies, inferred from each harmonic,
  refer to the epoch MJD 55482. In brackets are the statistical errors
  at the 68\% confidence level, calculated by means 
  of Monte Carlo simulations as in
  \citet{art-2008-hart-pat-etal}.} 
\label{table_prop}
\end{table}

\subsection{Pulsation analysis}

The data were folded in non-overlapping intervals of $\sim10-20$ s for
the first observation (where APP and BO fractional amplitudes are
$\sim30\%$), and $\sim100-500$ s for all others using the
ephemeris of \citet{art-2011-papit-etal} to search for both
APPs and BOs. This means that in some cases (when the signal is
sufficiently strong) multiple BO profiles are constructed for each
burst. Only pulse harmonics with SNR$>3.5$ were retained (i.e.,
harmonics whose amplitude was larger than 3.5 times the amplitude
standard deviation), giving less than 1 false pulse detection.

Each pulse profile was decomposed into the fundamental frequency and
 overtones to compute fractional amplitudes and phases (we were
able to identify up to five overtones in some intervals).  The pulse
phases of each profile were calculated for each harmonic and analyzed
separately. The method applied is described in
\citet{art-2010-pat-hart}, along with the procedure to calculate pulse
amplitudes and statistical errors. Phases were then fitted
with a Keplerian orbit, a linear and a parabolic term
representing the pulse frequency and its first time derivative.

\section{Discovery and properties of burst oscillations}
\label{bosc}

Burst oscillations were detected in all 231 bursts identified.
Fig. \ref{fracamp} shows the fractional amplitudes of the BOs compared
to those of the APPs. The fractional amplitudes of BOs are always of
the same order or larger than those of APPs, and always significantly
larger than expected if the modulations stem only from residual
persistent emission.  The detection of BOs can therefore be considered
secure.

Note that the APP contribution has not been subtracted when
calculating BO fractional amplitudes and phases. The measured BO
fractional amplitudes are comparable to or higher than the APP ones,
it can be seen from Eq. (11) of \citet{art-2005-watts-stroh-mark} and
Eq. (1) of \citet{art-2008-watts-pat-klis} that the correction is less
than a few percent.

The presence of timing noise, as seen in other AMXPs
\citep[]{art-2009-pat-wijn-klis} complicates timing analysis of APPs
and BOs.  The scatter observed in the oscillation phases has an
amplitude of 0.1-0.2 cycles within each ObsId for both BOs and
APPs. However adjacent BOs and APPs are always phase coincident and
phase locked to within $\sim 0.05$ cycles. In the first observation,
for example, which has the highest SNR, the phase difference between
the peak of the fundamental for APPs and BOs is $0.014\pm 0.005$
cycles ($5.1\pm 1.8$ deg). This coincidence of results is remarkable
given the strong timing noise, highlighting the fact
that BOs do indeed closely track APPs.

To check whether BO phases have the same temporal dependence as the
APPs, we fit a pulse frequency model plus its first derivative to the
the entire two data sets of BO and APP phases separately.  The pulse
frequency and derivative obtained are consistent with being the same
to within two standard deviations for both the fundamental frequency
and the first overtone (Tab. \ref{table_prop}).

Inspecting the behavior of the pulse frequency locally (comparing BO
and APP pulse frequencies in each individual ObsId), pulse frequency
is again consistent with being the same for the two data sets within
to two standard deviations. In the first observations, when the BO and
APP amplitudes are around 30\% (Fig.~\ref{powerspec}), the frequencies
of APPs (11.044881(2) Hz) and BOs (11.04493(7) Hz) are identical
within the errors ($\sim7\times10^{-5}$ Hz).

Comparing the properties of \sour{} to those of the other persistent
pulsars with BOs
\citep{art-2003-chak-morgan-etal,art-2003-stroh-mark-swank-zand,art-2010-alta-watts-etal,
  art-2011-riggio-etall}, \sour{} shows BOs in every burst, just like
SAX J1808.4-3658, XTE J1814-338 and IGR J17511-3057. \sour{} is most
similar to J1814: BOs are present throughout the bursts, BOs and APPs
are phase-locked very closely, BOs have strong harmonic content and no
measurable frequency drifts. J1808 and J17511, by contrast, have BOs
with weaker harmonic content, which do not persist throughout the
bursts, and display fast drifts in the rise. In terms of BO amplitude,
the persistent pulsars differ: J1808
has amplitudes of 3-5\% r.m.s., J17511 of 5-15\% and J1814 of
9-15\%. \sour{} has BO amplitude of 30\% r.m.s. in
the first burst, dropping to $\sim2\%$ thereafter. However, in all cases the BO amplitude
is comparable to (within a few percent) the APP amplitude.

\section{Burst oscillation models for \sour}
\label{theor}

We first show how we can exclude mode and Coriolis force confinement
models based on the coincidence of BO and APP frequencies and on slow
rotation.
\subsection{Global Modes}
\label{sec_modes}
Given a mode with azimuthal number $m$ and 
frequency $\nur$ in the rotating frame of the star,  
the frequency $\nuo$ an inertial observer would measure is
\begin{equation}
\nuo= m\nus + \nur
\end{equation}
the sign of $\nur$ being positive or negative, depending
on whether the mode is prograde or retrograde.

In the first, and best constrained burst (Section \ref{bosc}), $\nuo$
and $\nus$ were found to differ by no more than $10^{-4}$ Hz.
Excluding modes with $m>2$, which cannot explain the high fractional
amplitudes ($\sim 30\%$) observed in the first burst
\citep{art-2004-hey}, leaves us with two main possibilities.

The first is that we have modes with $|\nur| \sim \nus$ and m=0
(prograde) or m=2 (retrograde). Even though modes with frequencies
around 10 Hz, such as g-modes \citep{art-1998-cumm-bild} do exist, it
would require extreme fine tuning of parameters for $\nur$ to match
$\nus$ within $\sim 10^{-4}$ Hz. We are not aware of any mechanism
that could tie the frequency of a mode so closely to the spin.

The second option would be to have prograde modes with $m=1$ and
$\nur \sim 10^{-4}$ Hz. However the process that excites a mode must have a
time scale $\tau\sim 1/\nu$. In this case $\tau \sim 10^{4}$ s, which
is longer than the duration of the bursts.  We therefore conclude that
global modes cannot explain the presence of 
BOs in \sour{}.

\subsection{Coriolis force confinement}
\label{sec_coriolis}
For rapidly rotating stars, \citetalias{art-2002-spit-levin-ush}
showed that burning could be confined, due to the effect of the
Coriolis force, over a length scale of order the Rossby adjustment
radius \citep{book-1987-Pedlo}:

\begin{equation}
\Ro  =  \sqrt{gH}/4\pi\nus =  34\; \mathrm{km} 
\;M_{1.4}^{1/2}\;R_{10}^{-1}\;
H_{10}^{1/2} \;\nusten^{-1}
\end{equation} 
where $M_{1.4}$ is mass in units of 1.4 M$_\odot$, $R_{10}$ radius
in units of 10 km, $H_{10}$ the scale height of the burning fluid in units of
10 m \citepalias{art-2002-spit-levin-ush} and $\nusten = \nus/$(10 Hz).

For \sour{}, $\Ro$ significantly exceeds the stellar
radius: Coriolis force confinement is not effective.  So 
although this mechanism might be important for faster 
NSs (for $\nus \ge 200$ Hz, $\Ro \le 1.7$ km), it
cannot cause BOs in \sour.

\section{Magnetic confinement}
\label{sec_bfield} 

Having ruled out these two models, we now explore the possibility
that the magnetic field could confine the hot-spot.  We begin by
placing limits on the strength of the magnetic field for this source,
following the standard procedure outlined in
\cite{art-1999-psaltis-chak}.  For the magnetic field to be strong
enough to channel the accretion flow, and hence generate APPs, the
magnetospheric radius (the point at which the magnetic field disrupts
the disc) must exceed the stellar radius.  This gives a lower limit on
the magnetic field.  In addition, to avoid propeller effects (which
would inhibit accretion), the magnetospheric radius has to be lower
than the corotation radius.  This gives an upper limit on the magnetic
field.  Using the accretion rates derived in Section
\ref{subsec_light} we conclude that the magnetic field for J17480 is
in the range $2\times 10^8$ to $3\times 10^{10}$ G (see
\citet{art-1999-psaltis-chak} and
\citet{art-2005-ander-glamp-hask-watts} for a detailed explanation of
how to calculate the magnetic field).

We now need to estimate whether this field is strong enough to confine
the burning material in order to generate BOs. We start by considering
static fuel confinement. A mountain of accreted material induces a
pressure gradient within the fuel. \citet{art-1998-brow-bild} computed
the field that can statically compensate for such a gradient by
magnetic tension. Confinement within an area of radius $R_c$, at the
ignition column-depth $\sim 10^8$ g/cm$^2$ requires magnetic field

\begin{equation}
B \gtrsim 3\times 10^{10}~\mathrm{G} \left(\frac{R_c}{1~\mathrm{km}}\right)^{-1/2} 
\label{bb}
\end{equation}

Fuel confinement may however be ineffective due to MHD interchange or
``ballooning'' instabilities \citep{art-2001-lit-brow-rosn}, which
allow accreted fluid to escape the polar cap region and spread over
the NS surface.  Nonetheless, confinement might still be achieved
dynamically by the motion of the fluid itself.  This is because such
motion can induce a horizontal field component which can act to halt
further spreading.  This mechanism requires a weaker field

\begin{eqnarray}
B \gtrsim 4\times 10^9 ~\mathrm{G} \left(\frac{R_c}{1~\mathrm{km}}\right)^{-1}
\label{hs}
\end{eqnarray}
\citep{art-2009-heng-spit}.

Both of these estimates are compatible with the magnetic field
inferred for this source from its accretion properties.  We conclude
that magnetic confinement of the burning fluid is a viable mechanism
to explain the BOs in \sour.  Detailed theoretical studies will
however be needed to verify this possibility, since it depends on
poorly understood details of the interaction between the magnetic
field and the hydrodynamics of the burning ocean. That said,
confinement of burning material at the polar cap provides a natural
explanation for the phase-locking and coincidence of the APPs and BOs
\citep[see also][]{art-2007-lovel-kulk-roma} and would explain why the
amplitudes of the two sets of pulsations track each other as the
accretion footprint varies during the outburst (Fig. \ref{fracamp}).

\section{Conclusions and wider implications}
\label{conc}

We have analyzed the bursts of the slowly rotating X-ray pulsar
\sour{} and found oscillations at the same frequency as the NS
spin. We have shown that neither global modes nor a Coriolis force confined
hot-spot can explain the presence and frequency of these BOs.  We suggest
that the magnetic field could potentially provide the necessary confining force for
a hot-spot to persist.  The field needs to be at least $B\gtrsim 10^9$
G and this requirement is compatible with the constraints set by the
accretion process.  This model would neatly explain the phase-locking of
APPs and BOs, and the fact that the amplitudes of the two pulsations
track each other.

We noticed that among the other persistent AMXPs XTE J1814-338 is the
most similar one, showing almost constant frequency of the BOs,
harmonic content and remarkably phase
locking. \citet{art-2008-watts-pat-klis} excluded the possibility of
magnetic confinement, since using the estimates of
\citet{art-1998-brow-bild} (Eq. \ref{bb}) the field was not strong
enough to confine the fuel. However, they did not consider the case of
dynamical confinement when spreading is allowed (Eq. \ref{hs}). This
operates at lower fields which are compatible with the magnetic field
estimated for J1814.
 
%
This mechanism seem less likely for other AMXPs such as SAX
J1808.4-3658 and IGR J17511-3057, where the fields are lower
\citep{art-2009-hart-pat-etal} and there are frequency drifts in the
rising phase of bursts
\citep{art-2003-chak-morgan-etal,art-2010-alta-watts-etal}. That said,
magnetic confinement would offer an explanation for the presence of
higher harmonics in pulsar BOs, since the emitting area
would be bounded \citep[probably near the rotational
poles,][]{art-2002-muno-ozel-chakra}.


What about the non pulsating sources and the intermittent pulsars with
a weaker magnetic field? They do not show harmonic
content and their BOs show drifts during the decay of the
bursts. If magnetic confinement were responsible one would expect to
see APPs from these sources as well.  Unless the field can be boosted
temporarily to detectable levels only during the burst
\citep{art-2010-boutloukos}, a different mechanism may be required
\citep{art-2006-watts-stroh}.  If global modes or Coriolis confinement
are responsible, then the wider emitting surface involved, or the
involvement of higher colatitudes, would suppress the presence of
harmonics, in line with the observations
\citep{art-2002-muno-ozel-chakra, art-2003-stroh-mark-swank-zand}.

\acknowledgments 

We acknowledge support from NOVA (YC), an NWO Veni
Fellowship (AP), an EU Marie Curie Fellowship (BH), and an NWO
Rubicon Fellowship (ML). RW is supported by an ERC starting grant.

\bibliographystyle{apj}               

\end{document}